\documentclass[conference]{IEEEtran}
\usepackage{amsthm,amsmath,amsfonts}
\usepackage{amssymb}
\usepackage{array}
\usepackage[caption=false,font=normalsize,labelfont=sf,textfont=sf]{subfig}
\usepackage{textcomp}
\usepackage{stfloats}
\usepackage{url}
\usepackage[hyperfootnotes=false]{hyperref}
\usepackage{verbatim}
\usepackage{graphicx}
\usepackage{cite}
\usepackage{todonotes}
\usepackage{algorithm,algpseudocode}
\newtheorem{theorem}{Theorem}
\newtheorem{corollary}{Corollary}
\newtheorem{definition}{Definition}
\newtheorem{example}{Example}
\newtheorem{case}{Case}
\newtheorem{remark}{Remark}

\usepackage{tikz}
\usetikzlibrary{calc,trees,positioning,arrows,chains,shapes,  decorations.pathreplacing,decorations.pathmorphing,shapes,  matrix,shapes.symbols}
\usepackage{amssymb}

\usepackage{comment}



\def\I{{\mathcal I}}

\def\B{{\mathcal B}}

\def\C{{\mathbb C}}

\def\Q{{\mathbb Q}}
\def\R{{\mathbb R}}

\def\Z{{\mathbb Z}}

\def\z{{\bf z}}
\def\tr{{\rm Tr}}

\begin{document}

\title{Certified Real Eigenvalue Location}

\author{
\IEEEauthorblockN{Baran Solmaz}
\IEEEauthorblockA{\textit{Dept. of Computer Engineering} \\
\textit{Gebze Technical University}\\
Kocaeli, Turkey \\
b.solmaz2018@gtu.edu.tr}
\and
\IEEEauthorblockN{Tülay Ayyıldız\textsuperscript{1}}
\IEEEauthorblockA{\textit{Dept. of Computer Engineering} \\
\textit{Gebze Technical University}\\
Kocaeli, Turkey \\
tulayayyildiz@gtu.edu.tr}
}

\maketitle
\begin{abstract}
The location of real eigenvalues provides critical insights into the stability and resonance properties of physical systems. This paper presents a hybrid symbolic–numeric approach for \textit{certified real eigenvalue localization}. Our method combines Gershgorin disk analysis with Hermite matrix certification to compute certified intervals that enclose the real eigenvalues. These intervals can be further refined through bisection-like procedures to achieve the desired precision. The proposed approach delivers reliable interval certifications while preserving computational efficiency. The effectiveness of the framework is demonstrated through a concise, fully worked computational example.



\end{abstract}

\begin{IEEEkeywords}
Real Eigenvalue Location, Gershgorin Disks, Hermite Matrices, Real Root Isolation.
\end{IEEEkeywords}

\footnotetext[1]{ The corresponding author.}
\section{Introduction}\label{sec:Intro}
The linear eigenvalue problem is a fundamental concept in linear algebra, defined for a matrix $A \in \mathbb{R}^{n \times n}$ as finding scalars $\lambda \in \mathbb{C}$ and non-zero vectors $x \in \mathbb{R}^n$ such that $(A - \lambda I)x = 0$. Here, $\lambda$ is an eigenvalue and $x$ is its corresponding eigenvector \cite{meyer2000matrix}. This problem can be extended to a nonlinear form, where the goal is to solve $T(\lambda)x = 0$ for $x \neq 0$, with $T(\lambda)$ being an $n \times m$ matrix-valued function of the scalar parameter $\lambda$ \cite{guttel2017nonlinear}. When $T(\lambda)$ has a polynomial structure, i.e., $T(\lambda) = \sum_{i=0}^{k} \lambda^i A_i$ where $A_i \in \mathbb{C}^{n \times m}$, the nonlinear eigenvalue problem reduces to computing the roots of a univariate polynomial \cite{mackey2015polynomial}.

Linear and polynomial eigenvalue problems have arisen in many applications as they are or as an approximation of other, more general, nonlinear eigenvalue problems \cite{mehrmann2004nonlinear}. In many applications and iterative solution methods \cite{van2016designing} \cite{van2016nonlinear}, the location of the roots is crucial information. Specifically, real eigenvalues are important in assessing system stability, resonance, and bifurcation behavior, especially in structural dynamics and physical modeling.

As noted by G. H. Golub and H. A. van der Vorst (2000) \cite{GOLUB200035}, real eigenvalues often appear in symmetric or Hermitian matrix problems, where they play a central role in efficient numerical computation and the convergence properties of iterative solvers. They are indicative of well-posed problems and allow for the use of powerful divide-and-conquer strategies in large-scale matrix computations.

Similarly, Lin, Mottershead, and Ng (2020) \cite{lin2020state}, emphasize that in engineering applications, such as vibration analysis and modal testing, real eigenvalues are associated with undamped natural frequencies, which are physically observable and critical for system design and diagnostics. The sensitivity of real eigenvalues to structural changes also makes them essential in damage detection, control design, and model updating.

In this work, we consider the case where the eigenvalues are the roots of a polynomial, namely the characteristic polynomial. We leverage certified real root isolation techniques and combine them with eigenvalue bounds to obtain certified intervals that contain the real eigenvalues of a given matrix. The computation starts from known eigenvalue locations. 

So-called (generalized) Hermite matrices \cite{Parrilo06Lecture16} \cite{BasuPollackRoy2016} are used to certify real approximate solutions of multivariate polynomial systems over $\Q$, \cite{AYYILDIZAKOGLU2023101}. The approximate solutions are certified on a defined neighborhood. 
We propose a method that uses Hermite matrices to certify the presence of real eigenvalues within Gershgorin disks \cite{Ger31}.

First, in Section \ref{sec:Preliminaries}, we present the fundamental definitions and notation. Section \ref{sec:hermiteMatrices}
introduces the Hermite certification method for real root isolation for real polynomials. Then we explain how to construct the Hermite matrices without any knowledge of the roots or the eigenvalues, see Section \ref{sec:ConstructionOfHermite}. Using the Gershgorin Disks and Hermite matrices, we give the first intervals that contain only the real eigenvalues. We refine these intervals by using a second kind of Hermite matrices, this time defined on intervals on real line not as circles on the complex plane.


%
As the main part of this work, Section \ref{sec:eigenvalue-localization} presents our approach based on generalized Hermite matrices for eigenvalue localization within circles and intervals, culminating in algorithms that compute certified intervals containing real eigenvalues. The efficacy of our framework is then demonstrated through an illustrative computational example in Section~\ref{sec:example}. A Julia implementation of the algorithms accompanies this work and is available at \url{https://github.com/baransolmaz/Certified-Real-Eigenvalue-Location}.


As one of the most important parts of this study, Section \ref{sec:Analysis} represent the novelty of our approach and how it should not been considered as alternative to numerical eigenvalue solvers as it is a real eigenvalue certification tool. 
Finally, Section \ref{sec:comp_details} provides the computational details of our implementation.

\section{Preliminaries}\label{sec:Preliminaries}
In this section, we introduce the fundamental concepts we use for our method. The definitions of generalized Hermite matrices, Hermite Theorem, Gershgorin Disks, and La Budde's method for characteristic polynomial calculation will be introduced.

\subsection{Hermite Matrices}
We summarize the key concepts of (generalized) Hermite matrices for univariate polynomials, following the comprehensive treatments in \cite{BasuPollackRoy2016, AYYILDIZAKOGLU2023101}.
 
We assume the polynomial \(p(x) \in \R[x]\) corresponding to the companion matrix is square-free, i.e., \(\gcd(p, p') = 1\). The extension to polynomials with repeated roots is omitted here for brevity.

\begin{definition}\label{def:Hermite1}
Let $p(x) \in \R[x]$ be a degree n univariate polynomial with distinct roots $z_1,\ldots,z_n \in \C$, given in the standard basis $\mathcal{B}=\{1,x,\dots, x^{n-1},x^n \}$. Then the Hermite matrix of $p(x)$ with respect to the auxiliary polynomial $q(x) \in \R[x]$ is defined by
\begin{equation}\label{eqn:Hermite_mtxform}
	H_q(p):=V^T D_q V
\end{equation}
	\noindent
	where $V=[\z_{i}^{j-1}]_{i,j=1, \ldots, n}$ is the Vandermonde matrix of the roots with respect to the basis $\mathcal{B}$ and $D_q$ is a $n\times n$ diagonal matrix with $[D_q]_{ii}=q(z_i)$ for $i=1,\dots,n$.
\end{definition}
	
In general, the roots of $p(x)$ can have repetitions, so the Vandermonde matrix may have a rank deficiency. If we extend this definition, to reflect this fact, we obtain the following:
\begin{equation}
	[H_q(p)]_{i,j}=\sum_{k=1}^{n} (x^{i+j-2}q)(z_k)
\end{equation}
 Then using the Stickelberger Theorem, in \cite[Section 4.3.2]{BasuPollackRoy2016} it is shown that the following definition of Hermite matrices is equivalent to the Definition \ref{def:Hermite1}. This second definition implies that  the Hermite matrix has a Hankel structure and its entries are real numbers.
	
\begin{definition}\label{def:hermite_trace}
	Let $p(x),~ q(x)\in \R[x]$. The {\em Hermite matrix} of $\I:=<p>$ with respect to $q$ is 
	\begin{equation}\label{eqn:Hermite_trace}
	    \left[H_q(p)\right]_{i,j=1}^{n}:=\tr (M_{qx^{i+j-2}}),
	\end{equation} 
	where $M_f$ denotes  the matrix of the multiplication map $\mu_f:\R[x]/\I\rightarrow \R[x]/\I, $ $ \mu_f(g):= g\cdot f +\I$ in the basis $\mathcal{B}$. 
\end{definition}

\begin{remark}\label{rmk:Companion}
    In the univariate case, $M$ is the companion matrix of the given $p(x)$. \cite{BasuPollackRoy2016}
\end{remark}
Here the signature is defined as the following: 

\begin{definition}
Let \(A\) be a real and symmetric matrix, then the signature of \(A\) is
\[
\sigma(A) := (\#\text{ of pos. eigenvalues}) - (\#\text{ of neg. eigenvalues})
\]
\end{definition}

In 1856, Hermite \cite{hermite1856nombre} stated the following theorem.

\begin{theorem}[Hermite Theorem]\label{Thm:Hermite}
Let \(p(x),~ q(x) \in \mathbb{R}[x]\),
\[
\sigma(H_{q}(p)) = N_{+} - N_{-}.
\]
where \(N_{+} := \#\{x \in \mathbb{R} \mid p(x)=0 \text{ and } q(x)>0\}\) and \(N_{-} := \#\{x \in \mathbb{R} \mid p(x)=0 \text{ and } q(x)<0\}\)
\end{theorem}

Note that Hermite matrices are real and symmetric with a Hankel structure. Therefore, their eigenvalues are also real.

\subsection{Gershgorin Disks}
\label{subsec:gershgorin}

The Gershgorin Disks Theorem \cite{Ger31} provides a computationally efficient method to bound the eigenvalues of a complex square matrix, offering critical insights into spectral localization without explicit eigenvalue computation.

\begin{theorem}[Gershgorin Disks Theorem]\label{thm: Gershgorin}
Let $A = [a_{ij}] \in \mathbb{C}^{n \times n}$. Define the \textit{$i$-th Gershgorin disk} as
\[
D_i = \left\{ z \in \mathbb{C} : \lvert z - a_{ii} \rvert \leq R_i \right\}, \quad 
\text{where} \quad 
R_i = \sum_{\substack{j=1 \\ j \neq i}}^n \lvert a_{ij} \rvert.
\]
Then:
\begin{enumerate}
    \item Every eigenvalue of $A$ lies in $\bigcup_{i=1}^n D_i$.
    \item If $k$ disks form a connected component disjoint from the remaining disks, then exactly $k$ eigenvalues (counting multiplicity) lie in this component.
\end{enumerate}
\end{theorem}
\subsection{La Budde's Method for Characteristic Polynomial Computation}
\label{labudde}

La Budde's method \cite{rehman2011budde} provides a numerically stable approach for computing the characteristic polynomial of real matrices through a two-stage process. First, the input matrix $A \in \mathbb{R}^{n \times n}$ is reduced to upper Hessenberg form $H$ via orthogonal similarity transformations $H = Q^T A Q$, preserving the characteristic polynomial while improving numerical stability. Second, the characteristic polynomial coefficients are computed recursively using the recurrence relation:

\begin{align*}
p_0(\lambda) &= 1 \\
p_1(\lambda) &= \lambda - \alpha_1 \\
p_i(\lambda) &= (\lambda - \alpha_i)p_{i-1}(\lambda)\\ 
             &- \sum_{m=1}^{i-1} h_{i-m,i} \beta_i \cdots \beta_{i-m+1} p_{i-m-1}(\lambda )
\end{align*}

where $\alpha_i$ are diagonal elements, $\beta_i$ are sub-diagonal elements, and $h_{i-m,i}$ are off-diagonal elements of $H$. The method specializes to the Sturm sequence for symmetric matrices and reduces to the Summation Algorithm for diagonal matrices. Key advantages include preservation of coefficient conditioning through orthogonal transformations, exclusive use of real arithmetic, $\mathcal{O}(n^3)$ complexity, and superior accuracy for indefinite and non-symmetric matrices compared to eigenvalue-based approaches like MATLAB's \texttt{poly} function.

\section{Hermite Matrices and Real Root Isolation }
\label{sec:hermiteMatrices}
In this study, we only consider the univariate case since we approach the eigenvalues as roots of (characteristic) polynomials, which are univariate.  

Three choices of the auxiliary polynomial $q(x) \in \R[x]$ play important roles in our method. 

\begin{case}{$q(x)=1$} This is the base case for our computation. if $z_{1}, \ldots, z_{n}$ are the roots of $p(x)$, the entries of the Hermite matrix are the power sums of the elementary symmetric functions of the roots: 
 \begin{equation}\label{eq:H1f}
	 \left[ H_1(p)\right]_{i,j=1}^{n} = \sum_{k=1}^{n}z_{k}^{i+j-2}.
\end{equation} 

\end{case}

\begin{remark}
The signature of $H_1(p)$ gives the number of all real roots of $p(x)$:
    \begin{equation}\label{eq:H1_signature}
	 \sigma(H_1(p))=\#\{x \in \mathbb{R} \mid p(x)=0\}
\end{equation} 
\end{remark}
We obtain this result by Theorem \ref{Thm:Hermite} with $q(x) > 0$.
Let 
\begin{equation}\label{eqn:N0}
 N_{0} := \#\{x \in \mathbb{R} \mid p(x)=0 \text{ and } q(x)=0\}   
\end{equation}
then $H_1=N_0+N_+ +N_-$.

\begin{case}{$q(x)=(x-a)(x-b)$ for $a,b \in \R$.} We use this case for our computations on the interval 
$(a,b) \subset \R$.
\end{case}

\begin{theorem}\label{thm:case2}
Let $p(x)\in \R[x]$ and
$q(x)=(x-a)(x-b)$ for $a,b \in \R$,
if $\sigma(H_q(p)) = \sigma(H_1(p))$ then there are no real roots of $p(x)$ in the interval $[a,b]$.
\end{theorem}
\textbf{Proof:} 
First, note that $q(x) < 0$ when $x \in (a,b)$ and $q(x) \geq 0$ when $x \notin (a,b)$. 

If $\sigma(H_1) = \sigma(H_q)$, then $N_0+N_+ + N_- = N_+ - N_-$
by (\ref{eqn:N0}) and Theorem \ref{Thm:Hermite}. Since $N_0$ and $N_-$ are nonnegative integers, the only solution is $N_- = N_0 =0$. Therefore, there is no $x\in \R$ such that $p(x)=0$ in $[a,b]$ . 

Moreover, the contrapositive of this statement is equally useful, which is: if $\sigma(H_q(p)) \neq \sigma(H_1(p))$ then there is at least one real root in $[a,b]$.

\begin{case}{$q(x)=|x-c|^2-r^2$ for $c,r \in \R$}. Here we set $q(x)$ to define neighborhoods as circles with center at $c$ and radius $r$. 

\end{case}

\begin{theorem}\label{thm:case3}
Let $p(x)\in \R[x]$ and
$q(x)=|x-c|^2-r^2$ for $c,r \in \R$,
if $\sigma(H_q(p)) = \sigma(H_1(p))$ then there is no real root within the $r-$neighborhood of $c$.
\end{theorem}
\textbf{Proof:} Following the same logic as above, if $\sigma(H_1) = \sigma(H_q)$, then $N_- = N_0 =0$. Therefore,there is no $x\in \R$ and $p(x)=0$ which satisfies $|x-c|^2-r^2 \leq 0$. This implies there is no real root of $p(x)$ within the circle defined by the center at $c$ and radius $r$.

\section{Construction of Univariate Hermite Matrices}
\label{sec:ConstructionOfHermite}

 In this section, we mainly follow \cite{AyyildizSzantoS19} to construct the exact Hermite matrices of the characteristic polynomial $H_q(p)$ where $p(x) \in \R[x]$ and an auxiliary polynomial, $q(x) \in \R[x]$. 

In \cite{AyyildizSzantoS19}, the univariate Hermite matrix $H_q(p)$ is constructed using approximate roots of $p(x)$. However, here we construct the entries of the matrix without any knowledge of the roots. First, we construct the base case, $H_1(p)$ using the Newton-Girard formula \cite{cox1997ideals} to find power sums of elementary symmetric functions as Equation \ref{eq:H1f}. In this formula, we only use the coefficients of the characteristic polynomial.

Then we generalize it to any given $q$ using the Definition \ref{def:hermite_trace} of the Hermite matrices  and the Remark \ref{rmk:Companion}.

\begin{algorithm}[H]\label{alg:ConstructHermite}
\caption{Hermite Matrix Construction }
\label{alg:HermiteViaPowerSums}
\begin{algorithmic}[1]
\Require $p(x)$, $ q(x) \in \R[x]$ where $n=degree(p)$
\Ensure Hermite matrix $H_q(p)$
\For{$i = 1$ to $n$}
    \For{$j = 1$ to $n$}
        \State Compute $(i+j-1)$-th Power sum 
        $S[i + j - 1]$ using Newton-Girard formula
        \State $H_1[i, j] \gets S[i + j - 1]$
    \EndFor
\EndFor
    \State $M \gets$ the companion matrix of $p$
    \State  Compute $H_q \gets H_1 \cdot q(M)$

\State \Return $H_q$
\end{algorithmic}
\end{algorithm}

\textbf{Correctness Proof:} 
At Step 3,  we use (\ref{eq:H1f}) to compute $H_1$. In the Definition \ref{def:hermite_trace},  \(M\) is the multiplication matrix \(M_{x}\) of \(p\) defined on \(\R[x] / \langle p \rangle\) with the basis \( \B \). Therefore, as stated in Remark \ref{rmk:Companion} that \(M\) is the companion matrix corresponding to $p(x)$. Therefore, \(p(M) = 0\), \(M\) must be similar to a diagonal matrix \(D\) with diagonal entries \(z_{1}, \ldots, z_{n}\). More specifically,
\[
M = V^{-1} D V,
\]
\noindent
where \(V\) is the Vandermonde matrix with respect to the standard basis \( \B \) \cite{meyer2000matrix}. Since we assume that the polynomial \(p\) corresponding to the companion matrix is square-free, the Vandermonde matrix is invertible and the Step 4:
\begin{eqnarray*}
 H_1 \cdot q(M)
&=& V^T I V \cdot q(V^{-1} D V)\\
&=& V^T V V^{-1} D_q V\\
&=& V^T D_q V\\
&=& H_q
\end{eqnarray*}
Thus, the algorithm returns the Hermite matrix $H_q(p)$.


\section{Real Eigenvalue Localization Using Hermite-Gershgorin Hybrid Method}
\label{sec:eigenvalue-localization}
Now we can introduce our method to locate certified intervals that contain all the real eigenvalues of a given matrix $A\in \R^{n\times n}$. In fact, our method works for $A\in \C^{n\times n}$ as long as the characteristic polynomial has real coefficients. Such examples of complex matrices that are similar to real matrices, block diagonal matrices with conjugate pairs or Hermitian matrices. For the sake of simplicity, we continue with real matrices.


Given $A\in \R^{n\times n}$, the main steps of our approach are the following:
\begin{itemize}
    \item First we need to construct the characteristic polynomial of $A$. We use the La Budde's algorithm \cite{rehman2011budde},Section \ref{labudde}. This is an effective algorithm by avoiding any explicit computation of determinants.
    \item The Gershgorin Disks can be obtained directly from its definition given in the Theorem \ref{thm: Gershgorin}.
\end{itemize}

The following Corollary of the Theorem \ref{thm:case3} exhibits the connection between Gershgorin Disks and the Hermite theorem:

\begin{corollary}\label{cor:case3}
Let $A$ be an ${n\times n}$ matrix with its corresponding characteristic polynomial $p(x) \in \R[x]$   and $D_i$ for $i=1,\ldots n$ be the Gershgorin Disks of $A$ with corresponding centers at $a_{ii}$ with radius $R_i$. Let $q(x):=|x-a_{ii}|^2-R_i^2$ for each Gershgorin Disk, if 
\[
\sigma(H_q(p)) = \sigma(H_1(p))
\]
then there is no real eigenvalue of $A$ within the $i$-th Gershgorin Disk centered at $a_{ii}$ with radius $R_i$ for $i=1,\ldots,n$.
\end{corollary}

\begin{itemize}
    \item Using the Corollary \ref{cor:case3}, we can determine the Gershgorin Disks which contain at least one real eigenvalue.
    \item Now we have the region, as the union of the certified disks for real eigenvalues on the complex plane. Since real eigenvalues lie on the real line, the intersection of these disks on the real line defines the certified real intervals which contains the real eigenvalues.
    \item For each interval, we use the following Corollary for further refinement.
\end{itemize}

\begin{corollary}\label{cor:case2}
Let $A$ be an ${n\times n}$ matrix with its corresponding characteristic polynomial $p(x) \in\R[x]$ and $I:=[a,b] \subset \R$. Let $q(x):=(x-a)(x-b)$, if 
$$\sigma(H_q(p)) = \sigma(H_1(p))$$
then there is no real eigenvalue of $A$ within the interval $[a,b]$.
\end{corollary}

\begin{itemize}
    \item Let $I_i:=[a_i,b_i]\subset \R$ for $i=1,\ldots m$ be the certified intervals which obtained by the intersection of Gershgorin Disks with real eigenvalues and the real line.  Let $q(x):=(x-a_i)(x-b_i)$ for each interval $I_i$, if  
\(\sigma(H_q(p)) = \sigma(H_1(p))\)
then there is no real eigenvalue of $A$ within the $i$-th interval $[a_i,b_i]$ for $i=1,\ldots,n$

\item The computation of the signature can be done using and indirect approach. Since Hermite matrices only have real eigenvalues, Descartes Rule of Signs implies that the difference between the sign variations of $p(x)$ and $p(-x)$ gives the signature \cite{cox2005using}. Alternatively, $QR$ decomposition can be used here.

\end{itemize}

\begin{algorithm}[H] 
\caption{Certified Intervals}
\label{alg:CertifiedIntervals}
\begin{algorithmic}[1]
\Require $A\in \R^{n \times n}$
\Ensure Intervals $I_j$ for $j=1,\ldots,m$ for some $m\in \Z$.

    \State Compute the characteristic polynomial $p(x)$.
    \State Compute the Gershgorin Disks $D_i \subset \C$ for $i=1,\ldots,n$
    \State$H_1 \gets HermiteMatrixConstruction(p,1)$
    \State Compute the signature $\sigma(H_1)$
\For{$i = 1$ to $n$}
     \State Define $q_i \gets |x-a_{ii}|^2-R_i^2$
      \State $H_{q_i} \gets HermiteMatrixConstruction(p,q_i)$
    \If{$R_i == 0$}
        \State Add $a_{ii}$ as real eigenvalue
    \ElsIf{$\sigma(H_1) \neq \sigma(H_{q_i})$}
            \State Mark $D_i$ as containing real eigenvalue (certified disks)
    \EndIf
\EndFor
    \State Compute $I_i:=[a_i,b_i]$ for $i=1,\ldots m$ from the certified disks with the real line. 
\For{$i = 1$ to $m$}
    \State Define $q_i \gets (x-a_i)(x-b_i)$
    \State $H_{q_i} \gets HermiteMatrixConstruction(p,q_i)$
    \If{{$\sigma(H_1) \neq \sigma(H_{q_i})$}}
        \State Mark $I_i$ as containing real eigenvalue (certified interval)
    \EndIf
\EndFor
\State \Return Certified intervals with real eigenvalues
\end{algorithmic}
\end{algorithm}

\section{Illustration of The Method} \label{sec:example}
To demonstrate our certified real eigenvalue localization approach on a simple example, we provide a step-by-step illustration using a $5\times5$ matrix.

The example matrix is intentionally chosen to:
\begin{itemize}
    \item Contain both real and complex eigenvalues,
    \item Have overlapping Gershgorin disks,
    \item Demonstrate interval refinement.
\end{itemize}

Julia implementations of these algorithms can be found \url{https://github.com/baransolmaz/Certified-Real-Eigenvalue-Location}.

\begin{example}
Consider the matrix $A$ defined as:
\[
A = \begin{bmatrix}
    1.25 & 1 & 0.75 & 0.5 & 0.25 \\
     1 & 0 & 0 & 0 & 0 \\
    -1 & 1 & 0 & 0 & 0 \\
    0 & 0 & 1 & 3 & 0 \\
    0 & 0 & 0 & 0.5 & 5
\end{bmatrix}
\]
We demonstrate our certified real eigenvalue localization method for this matrix. We list the computation 

\begin{itemize}
       \item \textbf{Characteristic Polynomial }: Using the La Budde's algorithm, we compute:
    \[
         p(\lambda) =     \lambda^5 -9.25\lambda^4 +24.75\lambda^3 -17\lambda^2 - 0.625\lambda - 8.875
    \]
    \item \textbf{Hermite Matrix}: Via Newton-Girard power sums, we construct $H_1$:
    \[
        H_1 = \begin{bmatrix}
              5 &    9.25 & 36.063 & 155.64 & 706.88 \\
           9.25 &  36.063 & 155.64 & 706.88 & 3349.74 \\
         36.063 &  155.64 & 706.88 & 3349.74 & 16240.4 \\
         155.64 & 706.88 & 3349.74 & 16240.4 & 79751.8 \\
         706.88 & 3349.74 & 16240.4 & 79751.8 & 394523
        \end{bmatrix}
    \]
    Signature: $\sigma(H_1) = 3$
    \item \textbf{Companion Matrix}:
    \[
        C_p = \begin{bmatrix}
            0 & 0 & 0 & 0 & 8.875 \\
            1 & 0 & 0 & 0 & 0.625 \\
            0 & 1 & 0 & 0 & 17 \\
            0 & 0 & 1 & 0 & -24.75 \\
            0 & 0 & 0 & 1 &  9.25
            \end{bmatrix}
    \]
    \item \textbf{Gershgorin Disks} (Figure \ref{fig:gershgorin-disks}):
        \begin{align*}
            &D_1: \text{center } 1.25, \text{ radius } 2.5 \\
            &D_2: \text{center } 0, \text{ radius } 1  \\
            &D_3: \text{center } 0, \text{ radius } 2  \\
            &D_4: \text{center } 3, \text{ radius } 1  \\
            &D_5: \text{center } 5, \text{ radius } 0.5
        \end{align*}
        \begin{figure}[!t]
            \centering
            \includegraphics[width=\columnwidth]{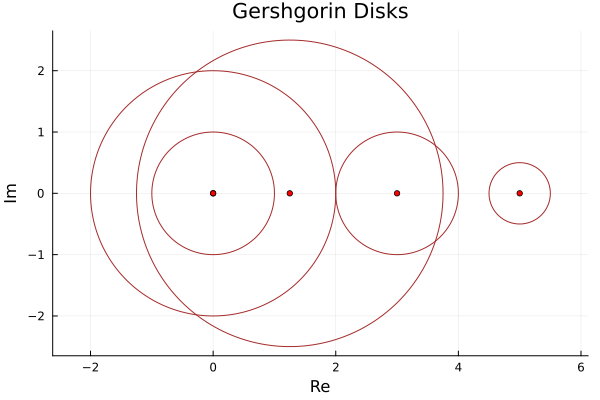}
            \caption{\label{fig:gershgorin-disks}Gershgorin Disks}
        \end{figure}

    \item \textbf{Disk Certification} (Figure \ref{fig:scanned-gershgorin-disks}):
        For each disk $D_i$ with center $c_i$ and radius $r_i$ :
    \begin{itemize}
        \item Define $q_i(x) = (x-c_i)^2 - r_i^2$
        \item Compute $H_{q_i} = H_1\cdot q_i(C_p)$
        \item Check if $\sigma(H_{q_i}) \neq \sigma(H_1) $:
            \begin{itemize}
                \item $D_1,D_3,D_4 \text{ and } D_5$ : Signature change $\xrightarrow{}$ Contain real eigenvalues
                \item Other disks: No signature change
                 \end{itemize}
    \end{itemize}
    \begin{figure}[!t]
        \centering
        \includegraphics[width=\columnwidth]{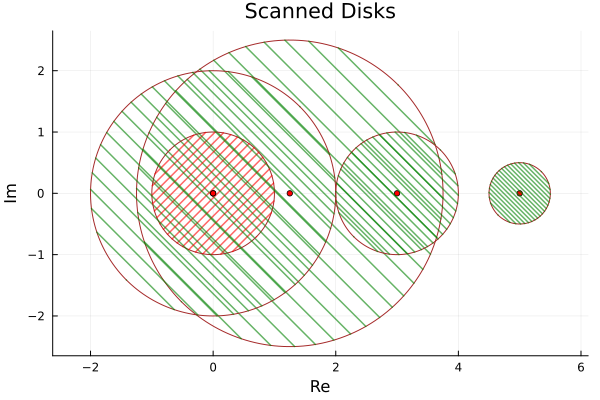}
        \caption{\label{fig:scanned-gershgorin-disks} Scanned Gershgorin Disks}
    \end{figure}
    \item \textbf{Interval Certification} (Figure \ref{fig:scanned-intervals}): Real candidate points: $ \{ -2,-1.25,-1,1,2,3.75,4,4.5,5.5 \}$ (boundaries and center of certified disks). Test intervals between consecutive points:
    \begin{itemize}
        \item For $[a,b]$, define $q(x)=(x-a)(x-b)$
        \item Compute $H_q=H_1 \cdot q(C_p)$
        \item Interval contains real eigenvalue if $\sigma(H_{q_i}) \neq \sigma(H_1)$, table \ref{tab:Interval certification results} :
    \end{itemize}
    \begin{table}[h]
    \centering
    \caption{Interval certification results}
    \label{tab:Interval certification results}
    \begin{tabular}{c|c|c}
        Interval & $\sigma(H_g)$ & Contains Real Eigenvalue \\
        \hline
        $\left[-2, -1.25\right]$    & 3 & No \\
        $\left[-1.25, -1\right]$    & 3 & No \\
        $\left[-1, 0\right]$        & 3 & No \\
        $\left[0, 1\right]$         & 3 & No \\
        $\left[1, 1.25\right]$      & 3 & No \\
        $\left[1.25, 2\right]$      & 1 & \textbf{Yes} \\
        $\left[2, 3\right]$         & 1 & \textbf{Yes} \\
        $\left[3, 3.75\right]$      & 3 & No \\
        $\left[3.75, 4\right]$      & 3 & No \\
        $\left[4,4.5\right]$        & 3 & No \\
        $\left[4.5,5\right]$        & 1 & \textbf{Yes} \\
        $\left[5,5.5\right]$        & 3 & No \\

    \end{tabular}
    \end{table}
    
    \begin{figure}[!t]
    \centering
    \includegraphics[width=\columnwidth]{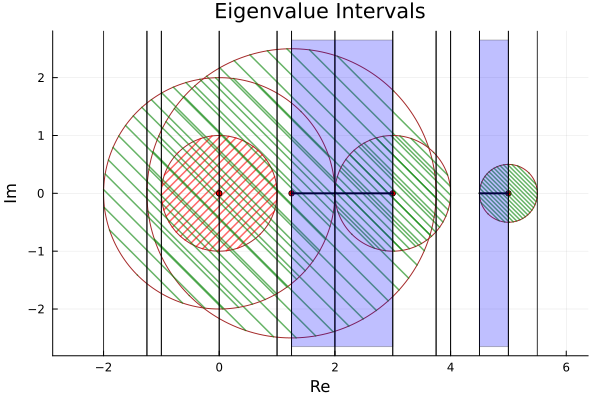}
    \caption{\label{fig:scanned-intervals} Intervals that has real eigenvalue}
    \end{figure}
    
\end{itemize}
\end{example}

\begin{table}[h]
\centering
\caption{Eigenvalues of matrix $M$}
\label{tab:eigenvalues}
\begin{tabular}{c|c}
$\lambda$ & Approximate Value \\ \hline
$\lambda_1$ & 4.997297881111628 \\
$\lambda_2$ & 2.934726733862349 \\
$\lambda_3$ & 1.732946083034579 \\
$\lambda_4$ & $-0.207485349004278 - 0.553312561675986i$ \\
$\lambda_5$ & $-0.207485349004278 + 0.553312561675986i$ \\
\end{tabular}
\end{table}

These intervals are guaranteed to contain at least one real eigenvalue each. The adjacency of $\mathcal{I}_1$ and $\mathcal{I}_2$ at $x=2$ does not imply eigenvalue duplication, as the signature test for each interval is independent.

Note that, the interval values in the Table \ref{tab:Interval certification results} can be refined further, we only show a partial results as we have a limited space here.

\section{Analysis of The Hermite-Gershgorin Hybrid Method}
\label{sec:Analysis}

The dominant cost of the entire computation is coming from the characteristic polynomial computation. We used La Budde's method as briefly explained in the Section \ref{labudde} as it is known for stability and its complexity is $O(n^3)$ \cite{rehman2011budde}.


We used a number of optimization techniques to reduce the computational cost. For example, for each input matrix, we compute $H_1$ and its signature only once, then reuse them to compute $H_q$ and its signature for different auxiliary polynomials $q$. Populating the matrix $H_1$ using the Newton-Girard formulas requires $O(n^3)$ operations. Although step 8 of Algorithm \ref{alg:ConstructHermite} involves matrix multiplication, its cost is reduced to $O(n^2)$ instead of the general $O(n^3)$ complexity. This improvement is due to the special structure of the companion matrix $M$, which has only one non-trivial column (containing the polynomial coefficients) and a sub-diagonal of 1's, with all other entries being zero. Additionally, the auxiliary polynomials $q$ are of degree two with a simple structure. Therefore, once $H_1$ is computed, computing $H_g$ requires only $O(n^2)$ operations. The total cost of Algorithm \ref{alg:ConstructHermite} is $O(n^3)$, which determines the overall complexity of our method for obtaining the initial certified intervals containing real eigenvalues.

When a bisection-based approach is used to refine the initial interval $[a,b] \subset \R$, the worst-case computational cost is
\[
    O\!\left(n^3 \log \frac{|a-b|}{\varepsilon}\right),
\]
where $\varepsilon$ is a user-defined precision parameter. Although this bound is rarely reached in practice, in the extreme case where the refinement produces $n$ disjoint certified intervals after the initial localization, the cost can be estimated as $O(n^4 log(|a-b| /\ \varepsilon))$. This worst-case scenario is, however, inherently parallelizable, and in practice the actual cost is typically much lower.

\begin{remark}

We emphasize that our method should not be regarded as an alternative to numerical eigenvalue solvers, which typically have a computational cost of $O(n^3)$ \cite{GOLUB200035}. Rather, our approach is designed specifically to certify the location of real eigenvalues, which play a crucial role in many applications, as discussed in the introduction. In contrast, numerical methods provide only approximate values that are not certified. For instance, if a computed eigenvalue has a small nonzero imaginary part, numerical methods cannot reliably determine whether the eigenvalue is truly real or complex. Our method, however, can rigorously certify whether an approximate eigenvalue is real.

\end{remark}

\section{Details of The Computation of The Certified Eigenvalue Localization Algorithm} \label{sec:comp_details}

In this section, we present a performance study of the proposed real root isolation algorithm, which integrates Gershgorin disk analysis, Hermite matrix signatures, and a divide-and-conquer interval refinement strategy.

\subsection{Experimental Setup}
The experiments were carried out in the Julia programming language using the \texttt{Plots}, \texttt{LinearAlgebra}, and \texttt{GenericSchur} packages. All computations were performed with arbitrary precision \texttt{BigFloat} arithmetic to minimize rounding errors in polynomial and matrix operations. The main steps of the algorithm are as follows:
\begin{enumerate}
    \item Compute the characteristic polynomial of the input matrix using La Budde's method.
    \item Obtain power sums of the roots via Newton--Girard formulas and construct the Hermite matrix.
    \item Determine the signature of the Hermite matrix, which encodes information on the number of real roots.
    \item Apply Gershgorin disk analysis to derive candidate regions that may contain eigenvalues.
    \item Refine candidate intervals using a binary divide-and-conquer scheme until the prescribed tolerance $\varepsilon$ is reached.
\end{enumerate}

\subsection{Interval Refinement}
The interval refinement procedure proceeds recursively. For each candidate interval $[a,b]$, the Hermite signature test is applied to detect sign changes in the root count:
\[
    g(x) = (x-aI)(x-bI),  \qquad    H_g = H_1 \, g(C_p),
\]
where $C_p$ is the companion matrix of the characteristic polynomial and $H_1$ is the Hermite matrix. A difference in signature between $H_1$ and $H_g$ certifies that the interval $[a,b]$ contains at least one real root.
If the interval width is greater than the tolerance $\varepsilon$, the interval is bisected; otherwise it is accepted as an isolating interval for an eigenvalue.

For performance evaluation, the following metrics are recorded:
\begin{itemize}
    \item Number of initial candidate intervals obtained from Gershgorin disks.
    \item Number of final isolating intervals containing real eigenvalues.
    \item Average and maximum width of the final intervals.
    \item Total execution time (in seconds).
\end{itemize}

\subsection{Results}

For the test matrix
\[
M =
\begin{bmatrix}
1.25 & 1 & 0.75 & 0.5 & 0.25 \\
1 & 0 & 0 & 0 & 0 \\
-1 & 1 & 0 & 0 & 0 \\
0 & 0 & 1 & 3 & 0 \\
0 & 0 & 0 & 0.5 & 5
\end{bmatrix},
\]
the algorithm successfully isolated three real eigenvalues with tolerance $\varepsilon = 10^{-7}$. The recursive interval refinement converged to narrow intervals containing the roots:
\[
\begin{aligned}
\lambda_1 &\in [1.7329460382, \; 1.7329461277], \\
\lambda_2 &\in [2.9347267151, \; 2.9347267747], \\
\lambda_3 &\in [4.9972978234, \; 4.9972978830].
\end{aligned}
\]
All interval widths are below the tolerance, ensuring certified isolation.

The execution time for the complete process was $11.293$ seconds on the performance machine. Table \ref{tab:performance} summarizes the performance.

\begin{table}[h!]
\centering
\caption{Performance for the $5 \times 5$ test matrix.}
\label{tab:performance}
\begin{tabular}{l c}
\hline
Number of candidate intervals (from Gershgorin disks) & 5 \\
Number of final eigenvalue intervals & 3 \\
Average final interval width & $\approx 6.0 \times 10^{-8}$ \\
Maximum final interval width & $8.95 \times 10^{-8}$ \\
Execution time & 11.293 s \\
\hline
\end{tabular}
\end{table}

\begin{remark}
It is important to note that if all matrix entries are represented as exact rational numbers (e.g., $0.5 = \tfrac{1}{2}$, $0.625 = \tfrac{5}{8}$) rather than as floating-point decimals, then the Faddeev-Leverrier method \cite{baer2021faddeev} produces the characteristic polynomial coefficients in exact rational arithmetic. In this case, the subsequent Hermite signature tests and interval refinement steps can be carried out without any rounding errors. 

As a consequence, the achievable tolerance improves significantly: 
the algorithm is able to certify isolating intervals down to 
$\varepsilon = 10^{-16}$, as demonstrated in our experiments.
\end{remark}




\section*{Acknowledgments}
The authors gratefully acknowledge the reviewers for their insightful and constructive comments. We also extend our appreciation to Zafeirakis Zefeirakopoulos for the valuable discussions that informed and enriched this study.\\

This research was partially supported by the Turkish Council of Science Foundation, project number 122F138.

\bibliographystyle{IEEEtran}
\bibliography{refs} 

\newpage

\vfill

\end{document}